\documentclass[preprint,prd,showpacs]{revtex4} 
\usepackage{graphicx}

\begin{document}

\title{A Fluid Analog Model for
Boundary Effects in Field Theory}

\author{ L.H. Ford}
 \email[Email: ]{ford@cosmos.phy.tufts.edu} 
 \affiliation{Institute of Cosmology  \\
Department of Physics and Astronomy\\ 
         Tufts University, Medford, MA 02155}
\author{N.F. Svaiter }
 \email[Email: ]{nfuxsvai@cbpf.br}
 \affiliation{Centro Brasileiro de Pesquisas Fisicas-CBPF \\
 Rua Dr. Xavier Sigaud 150 \\
Rio de Janeiro, RJ, 22290-180, Brazil}

\begin{abstract}
Quantum fluctuations in the density of a fluid with a linear phonon dispersion relation
are studied. In particular, we treat the changes in these fluctuations due to non-classical
states of phonons and to the presence of boundaries. These effects are analogous to similar
effects in relativistic quantum field theory, and we argue that the case of the fluid is
a useful analog model for effects in field theory. We further argue that the changes in the
mean squared density are in principle observable by light scattering experiments. 
\end{abstract}


\maketitle 
 
\baselineskip=14pt 

\section{Introduction}

It is well known that quantized sound waves, whose excitations are phonons, share several
properties with relativistic quantum fields, such as the electromagnetic field. This is
especially true when the phonon dispersion relation is approximately linear, which will
be assumed throughout this paper. There is a phononic analog
of the usual Casimir effect, but it tends to be quite small. For example, the force on two
parallel plates due to phonon zero point energy is smaller than that in the electromagnetic
case by the ratio of the speed of sound in the fluid to the speed of light~\cite{DLP}.
This ratio is typically of order $10^{-6}$. However, forces due to
classical stochastic sound fluctuations have been discussed recently by several 
authors~\cite{Larraza,Bschorr,SU02,RP05,Recati,Lamoreaux},
and can be larger.  Here we will study the local changes in density 
fluctuations of a fluid due either to the presence of boundaries or to changes in the
quantum state of the phonons. This is an analog of the effect of boundaries on the
quadratic expectation values of relativistic quantum fields, such as the mean squared
electric field. 
 In a related context, Unruh~\cite{Unruh} has shown that the velocity potential
$\phi$, of a moving fluid with velocity ${\bf v} = \nabla \phi$, satisfies 
the same equation as does a relativistic scalar field in a curved spacetime.
The present paper is an expanded version of Ref.~\cite{FS08}.
We begin in Sect.~\ref{sec:nobound} by reviewing the quantization of
sound waves in a fluid and the calculation of the density correlation function. We also
review recent work on the scattering of light by zero point fluctuations in a fluid.
In Sect.~\ref{sec:squeezed}, we consider the effects of a squeezed state of phonons on
the local density fluctuations. The effects of boundaries are treated in Sect.~\ref{sec:bound},
where several different geometries are discussed. Our results are summarized and discussed
in Sect.~\ref{sec:final}.

\section{Phonons and Density Fluctuations in a Fluid} \label{sec:nobound}
\subsection{Quantization and the Density Correlation Function}

 We consider the quantization of sound waves in a fluid with a
linear dispersion relation, $\Omega_q = c_S\, q$, where  $\Omega_q$
is the phonon angular frequency, $q$ is the magnitude of the wave
vector, and $c_S$ is the
speed of sound in the fluid. This should be a good approximation for
wavelengths much longer than the interatomic separation.  
Let $\rho_0$ be the mean mass density
of the fluid. Then the variation in density around this mean value is
represented by a quantum operator, $\hat{\rho}(\mathbf{x},t)$, which
may be expanded in terms of phonon annihilation and creation operators
as~\cite{LL-ST}
\begin{equation}
 \hat{\rho}(\mathbf{x},t) = \sum_{\mathbf{q}} (b_{\mathbf{q}}
 f_{\mathbf{q}} + b^\dagger_{\mathbf{q}}  f^*_{\mathbf{q}}) \,,
\end{equation}
where
\begin{equation}
 f_{\mathbf{q}} = \sqrt{\frac{\hbar \omega \rho_0}{2 V c_S^2}}
\; {\rm e}^{i(\mathbf{q} \cdot \mathbf{x} -\Omega_q\, t)} \,. 
\label{eq:mode_fnt}
\end{equation}
Here $V$ is a quantization volume. The normalization factor in 
Eq.~(\ref{eq:mode_fnt}) can be fixed by requiring that the zero point energy
of each mode be $\frac{1}{2} \hbar \Omega_q$ and using the expression for the 
energy density in a sound wave,
\begin{equation}
U = \frac{c_S^2}{\rho_0}\, {\hat{\rho}}^2\,.
\end{equation}
 In the limit in which $V
\rightarrow \infty$, we may write the density correlation function
as
\begin{equation}
\langle \hat{\rho}(\mathbf{x},t)\, \hat{\rho}(\mathbf{x}',t') \rangle
= \frac {\hbar \rho_0}{16 \pi^3 c_S^2}\, \int d^3q\, \Omega_q\,
 {\rm e}^{i(\mathbf{q} \cdot\Delta \mathbf{x} -\Omega_q\, \Delta t)}\,, 
                        \label{eq:rhorho}
\end{equation}
where $\Delta \mathbf{x} = \mathbf{x} - \mathbf{x}'$ and
$\Delta t = t-t'$. The integral may be evaluated to write the
coordinate space correlation function as
\begin{equation}
\langle \hat{\rho}(\mathbf{x},t)\, \hat{\rho}(\mathbf{x}',t') \rangle
= -\frac{\hbar \rho_0}{2 \pi^2 c_S}\; 
\frac{\Delta \mathbf{x}^2 +3 c_S^2 \Delta t^2}
{(\Delta \mathbf{x}^2 -3 c_S^2 \Delta t^2)^3}\,.
\end{equation}

This is of the same form as the correlation function for the time
derivative of a massless scalar field in relativistic quantum field
theory, 
$\langle \dot{\varphi}(\mathbf{x},t)\, \dot{\varphi}(\mathbf{x}',t')
\rangle$.
(This analogy has been noted previously in the literature. See, for example, 
Ref.~\cite{FF04}.)
Apart from a factor of $\rho_0$, these two quantities may be obtained
from one another by interchanging the speed of light $c$ and the speed of
sound $c_S$. If $c \rightarrow c_S$, then 
\begin{equation}
\langle \dot{\varphi}(\mathbf{x},t)\, \dot{\varphi}(\mathbf{x}',t')
\rangle \rightarrow \rho_0 \,
\langle \hat{\rho}(\mathbf{x},t)\, \hat{\rho}(\mathbf{x}',t')
\rangle\,.
\end{equation}

In the limit of equal times, the density correlation function becomes
\begin{equation}
\langle \hat{\rho}(\mathbf{x},t)\, \hat{\rho}(\mathbf{x}',t) \rangle
= -\frac{\hbar \rho_0}{2 \pi^2 c_S \, (\Delta \mathbf{x})^4} \,.
\label{eq:rho=t}
\end{equation}
Thus the density fluctuations increase as $|\Delta \mathbf{x}|$
decreases. Of course, the continuum description of the fluid and the
linear dispersion relation both fail as  $|\Delta \mathbf{x}|$
approaches the interatomic separation. Also note the minus sign in
Eq.~(\ref{eq:rho=t}). This implies that density fluctuations at
different locations at equal times are anticorrelated. By contrast,
when  $c_S|\Delta t| > |\Delta \mathbf{x}|$, then
$\langle \hat{\rho}(\mathbf{x},t)\, \hat{\rho}(\mathbf{x}',t) \rangle >0$
and the fluctuations are
positively correlated. This is complete analogy with the situation in
the relativistic theory. Fluctuations inside the lightcone can
propagate causally and tend to be positively correlated. Fluctuations 
in a fluid 
for which $c_S|\Delta t| < |\Delta \mathbf{x}|$ cannot have propagated
from one point to the other, and are anti-correlated. This can be
understood physically because an over density of fluid at one point
in space requires an under density at a nearby point.

\subsection{Light Scattering by Density Fluctuations}

In Ref.~\cite{FS09}, the cross section for the scattering of light by the zero point
 density fluctuations is computed for the case that the incident light
angular frequency is large compared to the typical phonon frequency.
 The result is
 \begin{equation}
\left(\frac{d\sigma}{d\Omega}\right)_{ZP} = \sqrt{2(1-\cos \theta)}\;
\frac{\hbar \omega^5\, {\cal V}\, \eta^4}{32
  \pi^2 c^5 c_S\rho_0}\;
(\mathbf{\hat{e}}_{{\mathbf k},\lambda} \cdot 
\mathbf{\hat{e}}_{{\mathbf k}',\lambda'})^2    \,, 
\label{eq:cross_ZP2}
\end{equation}
where $\theta$ is the scattering angle, ${\cal V}$ is the scattering
volume, and $\eta$ is the mean index of refraction of the fluid.
In addition, $\mathbf{\hat{e}}_{{\mathbf k},\lambda}$ and 
$\mathbf{\hat{e}}_{{\mathbf k}',\lambda'}$ are the initial and final polarization
vectors, respectively. The $\omega^5$ dependence
of the scattering cross section can be viewed as the product of the
$\omega^4$ dependence of Rayleigh-Brillouin scattering and one power
 of $\omega$ coming from the spectrum of
zero point fluctuations in the fluid. The factor of $\eta^4$ represents
the influence of the fluid on light propagation before and after the
scattering process, and arises as a product of a factor of $\eta$ in the
incident flux and a factor of $\eta^3$ in the density of final states~\cite{FS09}.  
Because light travels through 
the fluid at speeds much greater than the sound speed, light
scattering  reveals a nearly static distribution of density
fluctuations. Thus we can regard Eq.~(\ref{eq:cross_ZP2}) as a
probe of the fluctuations described by Eq.~(\ref{eq:rho=t}).
The scattering by  zero point fluctuations is inelastic, with
the creation of a phonon. Thus, the scattering described by
Eq.~(\ref{eq:cross_ZP2}) is strictly Brillouin rather than Rayleigh scattering.

This scattering by zero point density fluctuations should be compared to
the effects of thermal density fluctuations. The ratio of the zero point to the
thermal scattering for the Stokes line may be expressed as
 \begin{equation}
R \equiv \frac{(d\sigma/d\Omega)_{ZP}}{(d\sigma/d\Omega)_{TB}} =
 \sqrt{2(1-\cos \theta)}\,
\left(\frac{\hbar \omega}{2 k_B T}\right)\,
\left(\frac{c_S}{c}\right)\, \eta^4\, \left[
 \rho_0\, \left(\frac{\partial \epsilon}{\partial
  \rho_0}\right)_S \right]^{-2}\,.
\end{equation}   
The index of refraction, $\eta$, and the quantity 
$ \rho_0\, \left({\partial \epsilon}/{\partial  \rho_0}\right)_S$,
which involves a derivative of the fluid dielectric function with respect
to density at constant entropy
are both of order unity. Hence $R$ is primarily determined by the ratio of the
photon energy to the thermal energy, and the ratio of the speed of
sound to the speed of light.

Note that the zero point scattering cross section, Eq.~(\ref{eq:cross_ZP2}), is the
sole cross section at zero temperature. At finite temperature, Stokes line cross
section (describing the process in which a phonon is emitted) is modified by
the factor
\begin{equation}
\langle n_{\bf q} \rangle +1 = \frac{1}{{\rm e}^{\hbar \Omega_q/k T} -1} +1 \,,
\end{equation}
where $\langle n_{\bf q} \rangle$ is the mean number of phonons in mode ${\bf q}$,
and $\hbar \Omega_q$ is the phonon energy.
In the low temperature limit, $k T \ll \hbar \Omega_q$, this correction factor 
goes to unity, giving the zero point result. In the high
temperature limit,  $k T \gg \hbar \Omega_q$, it becomes
\begin{equation}
\langle n_{\bf q} \rangle +1 \sim \frac{kT}{ \hbar \Omega_q} + \frac {1}{2}
+O(1/T) \,.
\end{equation}
The leading term is the usual high temperature limit. The next term is the zero
point effect, giving rise to a contribution to the cross section proportional
to $\omega^5$. More precisely, it is $1/2$ of the zero point effect, the other
$1/2$ having been canceled by the thermal correction. Our view is that zero
point fluctuations are always present at all temperatures, but in this case
the thermal correction partially masks the zero point effect. However, the
half which remains is potentially observable. For experiments in the high
temperature limit, the $\omega^5$ part of the cross section is given by $1/2$
of the right hand side of Eq.~(\ref{eq:cross_ZP2}). 

Some numerical estimates for various fluids are given in Ref.~\cite{FS09}
for violet light with a wavelength of $\lambda = 350nm$.
For the case of liquid neon, $R \approx 0.13$, so that about $13\%$
of the Stokes line is due to zero point motion effects~\cite{footnote}, which might be
detectable experimentally. Even in the case of water at room temperature,
$R \approx 0.004$. Although small in absolute terms, this is surprizingly
large for a macroscopic quantum effect at room temperature. 

It is interesting to note that if one were to look only at the total Brillouin
cross section (Stokes plus anti-Stokes), the zero point effect would be masked
at high temperatures. The anti-Stokes line describes phonon
absorption, so in the limit that $\omega \gg \Omega_q$, its cross section is
of the same form as that for the Stokes line, but its thermal correction factor
is $\langle n_{\bf q} \rangle$. The total cross section from both lines
has a factor of
\begin{equation}
2\,\langle n_{\bf q} \rangle +1 = 
\coth \left(\frac{\hbar \Omega_q}{2kT}\right) \sim 
 \frac{2kT}{\hbar \Omega_q} +O(1/T)\,, \quad k T \gg \hbar \Omega_q\,.
\label{eq:total}
\end{equation} 
Here the thermal part completely masks the zero point part, leaving
a residue of order $1/T$. The same masking effect also occurs for the energy of a
collection of harmonic oscillators, which is proportional to the quantity in
Eq.~(\ref{eq:total}). The thermal effect on scattering is often described by a
structure factor. See, for example, Ref~\cite{Fetter}. The hyperbolic cotangent
form of the srtucture factor, corresponding to Eq.(\ref{eq:total}), was calculated
in Ref.~\cite{FF56}. 

In the remainder of this paper, we will discuss modifications to the local
density fluctuations due to the phonon state or to boundaries. These modifications
are at least in principle observable through changes in the scattering cross
section, Eq.~(\ref{eq:cross_ZP2}).

\section{Squeezed States of Phonons}\label{sec:squeezed}

Here we consider the case where the phonon field is not in the vacuum state,
but rather a squeezed state. The squeezed states are a two complex parameter family
of states in which the quantum uncertainty in one variable can be reduced with a 
corresponding increase in the uncertainty of the conjugate variable. See, for example,
Refs.~\cite{Caves,GC08} for a detailed treatment of the properties of the squeezed states.
 We will focus attention on the case of the squeezed vacuum 
states $|\zeta\rangle$ for a single mode, labeled by a single complex squeeze parameter
\begin{equation}
\zeta = r\,{\rm e}^{i\delta} \,,
\end{equation}
and defined by 
\begin{equation}
|\zeta\rangle = S(\zeta)\; |0\rangle\,.
\end{equation}
Here 
\begin{equation}
S(\zeta) ={\rm e}^{\frac{1}{2}[\zeta^* a^2 - \zeta (a^\dagger)^2]}
\end{equation}
is the squeeze operator and $a$ and $a^\dagger$ are phonon annihilation and creation 
operators for the selected mode. 
This set of states is of special interest because they are the states generated by
quantum particle creation processes, and they can exhibit local negative energy
densities. (See, for example, Refs.~\cite{KF93,BFR02}) 

Consider the shift in the mean squared density fluctuations between
the given state and the vacuum
\begin{equation}
\langle \hat{\rho}^2 \rangle_R = \langle \zeta|\hat{\rho}^2|\zeta \rangle
- \langle 0|\hat{\rho}^2|0 \rangle \,,
\end{equation}
the ``renormalized'' mean squared density fluctuation. The result for this quantity
in a single mode squeezed vacuum state for a plane wave in the $z$-direction is
\begin{equation}
\langle \hat{\rho}^2 \rangle_R = 
\frac{ \hbar\, \omega\,\rho_0}{c_S^2 V}\, \sinh r \;\left\{ \sinh r
-\cosh r\, \cos[2(kz-\omega t) + \delta] \right\}\,.  \label{eq:rho_sq}
\end{equation}
Here we have used the identities
\begin{equation}
S^{\dagger}(\zeta)\,a\,S(\zeta)=
a\,\cosh r-a^{\dagger}e^{i\delta}\sinh r \,,
\end{equation}
and
\begin{equation}
S^{\dagger}(\zeta)\,a^{\dagger}\,S(\zeta)=
a^{\dagger}\,\cosh r-ae^{-i\delta}\sinh r \,.
\end{equation}
Note that this quantity can be either positive or negative, but its time or space average
is positive. The suppression of the local density fluctuations in a squeezed state
is analogous to the creation of negative energy densities for a massless, relativistic
field. (Compare Eq.~(\ref{eq:rho_sq}) with Eq.~(48) and Fig.~8 in Ref.~\cite{BFR02}.)

\section{Boundaries}\label{sec:bound}

If we introduce an impenetrable boundary into the fluid, the phonon field will
satisfy Neumann boundary conditions
\begin{equation}
{\bf \hat{n} \cdot \nabla}  \delta \rho =0   \label{eq:BC}
\end{equation}
as a consequence of the impenetrability. Thus there will be a Casimir force on the
boundaries which is analogous to the Casimir force produced by electromagnetic
vacuum effects. For example, consider two parallel plates, which will experience
an attractive force per unit area of
\begin{equation}
\frac{F}{A} = \frac{\hbar \, c_S\, \pi^2}{480\, a^4}\,,    \label{eq:force}
\end{equation}
which is smaller than the electromagnetic case for perfect plates by a factor of
${c_S}/(2c)$, and is thus quite small in any realistic situation. 

Henceforth, we consider the local effect of boundaries on mean squared density
 fluctuations, and now define $\langle \hat{\rho}^2 \rangle_R$ to be the change due to
the presence of the boundary. This quantity is of interest both as an analog model for 
the effects of boundaries in quantum field theory, and in its own right. The shifts in
density fluctuations are at least in principle observable in light scattering 
experiments.

Our interest in the phononic analog model is inspired by the fact that
the study of boundary effects
in quantum field theory is an active area of research, and has given rise to
some recent controversies in the literature~\cite{Jaffe,MCW06}.
 One question is the nature of the physical cutoff
which prevents singularities at the boundary. 
An example of the subtleties is afforded by the mean
squared electric and magnetic fields near a dielectric interface. When the
material is a perfect conductor, these quantities are proportional to
$z^{-4}$, where $z$ is the distance to the interface. Specifically, in 
Lorentz-Heaviside units their asymptotic forms are
\begin{equation}
\left\langle E^{2}\right\rangle \sim \frac{3 \hbar c}{16\pi ^{2}}\frac{1}{z^{4}}
\label{eq:E2}
\end{equation}
and
\begin{equation}
\left\langle B^{2}\right\rangle \sim -\frac{3 \hbar c}{16\pi ^{2}}\frac{1}{z^{4}}\,.
\label{eq:B2}
\end{equation}
 One might expect
that a realistic frequency dependent dielectric function would remove
this singularity, but this is not the case. Instead one finds~\cite{SF05} that
\begin{equation}
\left\langle E^{2}\right\rangle \sim \frac{\sqrt{2}\,\hbar \omega _{p}}{32\pi } 
\frac{1}{z^{3}}
\end{equation}
and
\begin{equation}
\left\langle B^{2}\right\rangle \sim -\frac{5\, \hbar\omega _{p}^{2}}{96\pi c }
\frac{1}{z^{2}} \,,
\end{equation}
where $\omega_p$ is the plasma frequency of the material.
Thus some physical cutoff other than dispersion is required. For realistic
materials, it is likely to be surface roughness, but fluctuations in the
position of the boundary can also serve as a cutoff~\cite{FS98}.  
In a fluid, there is always a physical cutoff at the
interatomic separation, 

In the remainder of this paper, we will analyze 
$\langle \hat{\rho}^2 \rangle_R$ in different geometries.

\subsection{One or Two Parallel Plane Boundaries}

In both of these case, the renormalized density two-point function may be constructed
by the method of images. First consider the case of a single plate located at $z=0$.
Let $G_0$ denote the density correlation function in the absence of a boundary.
The two-point function which satisfies the boundary condition Eq.~(\ref{eq:BC}) 
on this boundary is
\begin{equation}
G = G_0(\Delta t,\Delta {\bf x}_T,z-z') + G_0(\Delta t,\Delta {\bf x}_T,z+z')\,
\end{equation}
where ${\bf x}_T$ is in the direction transverse to the plate. The renormalized 
two-point function is
\begin{equation}
 G_R = G_0(\Delta t,\Delta {\bf x}_T,z+z')\,.
\end{equation}
The resulting shift in the mean squared density is
\begin{equation}
\langle \hat{\rho}^2 \rangle_R = G_R\biggr|_{{\bf x}={\bf x}'} =
 -\frac{\hbar\, \rho_0}{32 \pi^2\, c_S\, z^4} < 0
\end{equation}
where $z$ is the distance to the boundary. For the case of two parallel planes, the
correlation function is given by an infinite image sum:
\begin{equation}
G = \sum_{n=-\infty}^{\infty} \left[G_0(\Delta t,\Delta {\bf x}_T,z-z'-2an) + 
G_0(\Delta t,\Delta {\bf x}_T,z+z'-2an) \right]\,,
\end{equation}
where $a$ is the plate separation. If we use the identity
\begin{equation}
\sum_{n=-\infty}^{\infty} \frac{1}{(n-x)^4} =
 \frac{1}{6}\,\frac{d^2}{dx^2} \sum_{n=-\infty}^{\infty} \frac{1}{(n-x)^2} =
 \frac{\pi^2}{6}\,\frac{d^2}{dx^2} \sum_{n=-\infty}^{\infty} \csc^2(\pi x)\,,
\end{equation}
we can obtain the result
\begin{equation}
\langle   \hat{\rho}^2 \rangle_R = -\frac{ \hbar\, \rho_0}{96 \, c_S\, a^4}\;
\left[ \frac{1}{15} +
\frac{3 - 2 \sin^2(\pi z/a)}{ \sin^4(\pi z/a)} \right]\,,
\end{equation}
where $z$ is the distance to one boundary.  
Note that $\langle \hat{\rho}^2 \rangle_R$ for both of these cases is negative 
everywhere. In the absence of a physical
cutoff, both of these expressions diverge as $z^{-4}$ near the boundaries, just as 
do the squared electric and magnetic fields near a perfectly reflecting plane.
In contrast to the force between two plates, Eq.~(\ref{eq:force}), the shift in
mean squared density is inversely proportional to the speed of sound,
$\langle   \hat{\rho}^2 \rangle_R \propto 1/c_S$. This is a general feature 
of all shifts due to boundaries.

\subsection{A Three-Dimensional Torus}

Here we consider a rectangular box with periodic boundary conditions in all three
spatial directions, with periodicity lengths $L_1$, $L_2$ and $L_3$. Thus the 
three-dimensional
space has the topology of $S^1\times S^1 \times S^1$. This is closely related
to the geometry of a waveguide, where the fluctuations of a relativistic scalar field
were discussed by Rodrigues and Svaiter~\cite{RS03}.
 As in the parallel plane case, an image
sum method may be employed to write
\begin{equation}
G = \sum_{\ell,m,n=-\infty}^{\infty} G_0(\Delta t,x-x'+\ell L_1,y-y'+m L_2,z-z'+n L_3) \,.
\end{equation}
This leads to the result
\begin{equation}
\langle  \hat{\rho}^2 \rangle_R = -\frac{ \hbar\, \rho_0}{2 \pi^2\, c_S}\;
{\sum_{\ell,m,n}}' \frac{1}{(\ell^2 L_1^2 +m^2 L_2^2 +n^2 L_3^2)^2}\,.
\end{equation}
Here the prime on the summation indices denotes that the $\ell=m=n=0$ term is
omitted. In this case, $\langle \hat{\rho}^2 \rangle_R$ is a negative constant.

\subsection{A Wedge}

Consider two intersecting plane which are at an angle of $\alpha$ with respect to each
other. Now consider a point inside of this wedge which is located at polar coordinates
$(r,\theta)$, where $r$ is the distance to the intersection line and $\theta < \alpha$.
This geometry was treated for the relativistic case by Candelas and Deutsch~\cite{CD79},
whose  Eq.~(5.39) yields
\begin{equation}
\langle \dot{\varphi}^2 \rangle = 
\lim_{{\bf x}'\rightarrow {\bf x}} \frac{\partial^2}{\partial t^2}\, G(x,x')
= \frac{c^2}{3 r^2}\, \lim_{\theta' \rightarrow \theta} \left(1 + 
\frac{\partial^2}{\partial \theta^2} \right) \, G_R(\theta,\theta') \,.
\label{eq:dotphisq}
\end{equation}
Here $G_R(\theta,\theta') = G(\theta,\theta') - G_0(\theta,\theta')$, where
\begin{equation}
G_0(\theta,\theta') = \frac{\hbar}{16 \pi^2 c^3 r^2}\, 
\csc^2 \left(\frac{\theta-\theta'}{2}\right)   \label{eq:G0}
\end{equation}
is the empty space two-point function, and 
\begin{equation}
G(\theta,\theta') = \frac{\hbar}{16 \alpha^2 c^3 r^2}\,
 \left\{\csc^2 \left[\frac{\pi(\theta-\theta')}{2\alpha}\right] +
 \csc^2 \left[\frac{\pi(\theta+\theta')}{2\alpha}\right] \right\} \label{eq:Gtotal}
\end{equation}
is the two-point function in the presence of the wedge. 
 
We may combine these results to find for the phononic case
\begin{eqnarray}
\langle  \hat{\rho}^2 \rangle_R &=& 
-\frac{ \hbar\, \rho_0}{1440 \pi^2\, c_S\, r^4\, \sin^4(\pi \theta/\alpha)} 
\nonumber \\ &\times&
\left\{(\pi-\alpha)(\pi+\alpha)  \sin^2(\pi \theta/\alpha) [(\pi^2 +11\alpha^2) 
 \sin^2(\pi \theta/\alpha) -30\pi^2] +45\pi^4 \right\}\,.
\end{eqnarray}
Again, this quantity is negative everywhere.

\subsection{A Cosmic String}

As is well known, the space surrounding a cosmic string is a conical space
with a deficit angle $\alpha < 2\pi$. Quantum field theory in this conical space
has been discussed by many authors, beginning with Helliwell and Konkowski~\cite{HK86},
and is similar to the wedge problem discussed above. Eqs.~(\ref{eq:dotphisq}) 
and (\ref{eq:G0}) hold for the cosmic
string as well as the wedge. However, Eq.~(\ref{eq:Gtotal}) is replaced by 
\begin{equation}
G(\theta,\theta') = \frac{\hbar}{4 \alpha^2 c^3 r^2}\,
 \csc^2 \left[\frac{\pi(\theta-\theta')}{\alpha}\right] \,,
\end{equation}
which is equivalent to Eqs.~(15) and (16) in Ref.~\cite{HK86}.   
At a distance $r$ from the apex, we find
\begin{equation}
\langle  \hat{\rho}^2 \rangle_R = 
-\frac{ \hbar\, \rho_0}{1440 \pi^2\, c_S \, \alpha^4\, r^4}\;(2\pi-\alpha)(2\pi+\alpha)
(11\alpha^2 +4\pi^2) \,,
\end{equation}
which is also negative everywhere provided that $\alpha < 2\pi$.

\subsection{Near the Focus of a Parabolic Mirror}

\begin{figure}
\begin{center}
     \scalebox{0.6}{ 
\includegraphics{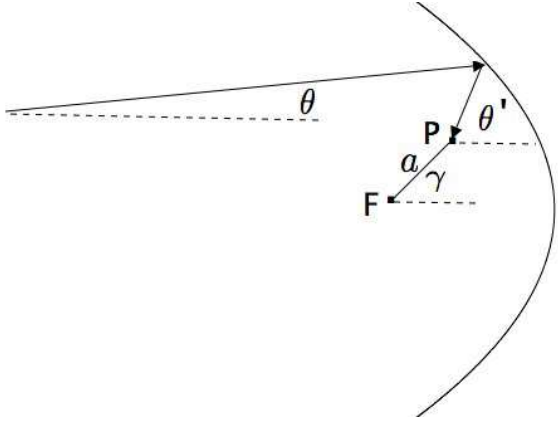}}
\end{center}
\caption{The geometry of rays reflecting from a parabolic mirror is illustrated.
 An incoming ray at an angle of $\theta$ reflects at an angle of
$\theta'$ to reach the point $P$, which is a distance $a$ from the focus $F$, 
and at an angle of $\gamma$.}
\label{fig:para}
\end{figure}

The quantization of the electromagnetic field in the presence of a parabolic mirror
was discussed by us in Refs.~\cite{FS00,FS02}, where a geometric optics approximation
was employed to find the mean squared fields near the focus. This treatment lead to the 
result that these quantities are singular at the focus, diverging as an inverse power
of the distance $a$ to the focus. This result holds both for parabolic cylinders and for
parabolas of revolution, and basically arises from the interference term of multiply
reflected rays with nearly the same optical path length. The geometry is illustrated in
Fig.~\ref{fig:para}. An incoming ray at an angle of $\theta$ reflects at an angle of
$\theta'$ to reach the point $P$, which is a distance $a$ from the focus $F$, 
as illustrated.
The distance from the focus to the mirror itself is $b/2 \gg a$. 

\begin{figure}
\begin{center}
 \scalebox{0.6}{ \includegraphics{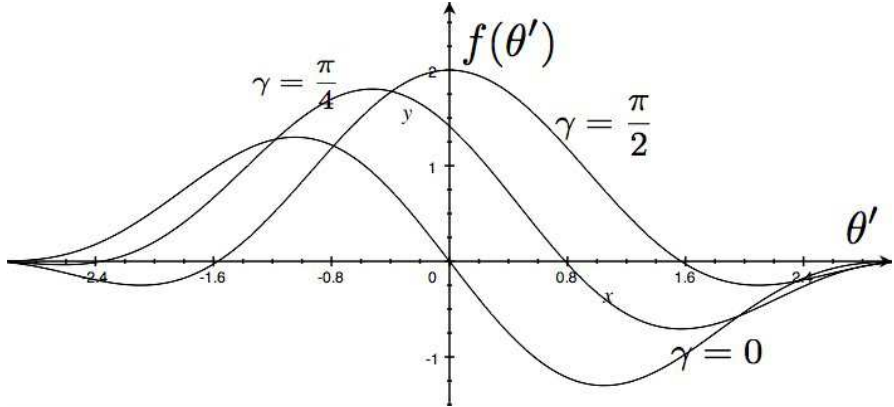}}
\end{center}
\caption{The function $f(\theta')$ is plotted for various values of $\gamma$.
This function relates the angle of the incident ray, $\theta$, to the angle
of the reflected ray, $\theta'$, through the relation 
$\theta = (a/b)\, f(\theta')$. }
\label{fig:f}
\end{figure}

The relation between $\theta$ and $\theta'$ is given by
\begin{equation}
\theta = \frac{a}{b} \, f(\theta') \,,
\end{equation}
where 
\begin{equation}
f(\theta') = -\frac{\sin^2 \theta' \, \sin (\theta'-\gamma)}{(1 - \cos \theta' )}
= -(1+\cos \theta')  \, \sin (\theta'-\gamma) \,. \label{eq:theta}
\end{equation} 
Note that $\theta$ is defined somewhat differently than in Refs.~\cite{FS00,FS02},
so that $f(\theta)$ now has the opposite sign.
There will be multiply reflected rays whenever different values of $\theta'$
are associated with the same value of $f$.
The function $f(\theta')$ is plotted in Fig.~\ref{fig:f} for various values of 
$\gamma$. We can see from these plots that in general there can be up to
four reflected angles $\theta'$ for a given incident angle $\theta$.
However, if the mirror size $\theta_0$ is restricted to be less than $2 \pi/3$,
then there will never be more than two values of $\theta'$ for a given $\theta$.
Throughout this paper, we will assume $\theta_0 < 2 \pi/3$, and hence have
at most two reflected rays for a given incident ray. The two reflected rays will
occur at $\theta'=\alpha$ and $\theta'=\beta$, where
\begin{equation}
f(\alpha) = f(\beta) \, .
\end{equation}
 
\begin{figure}
\begin{center}
     \scalebox{0.6}{ 
\includegraphics{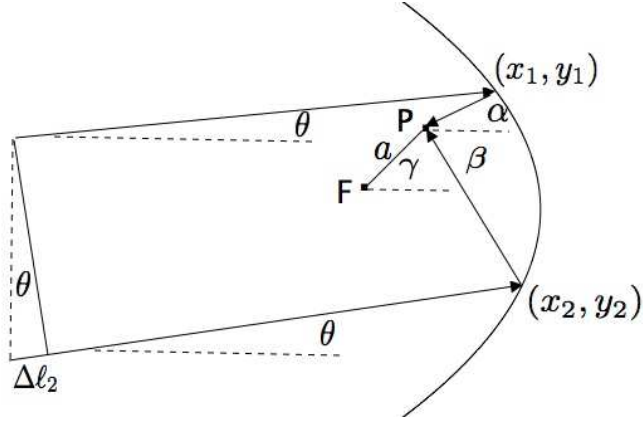}}
\end{center}
\caption{Two rays reflecting from a parabolic mirror to the point $P$ are illustrated.
The first arrives at an angle of $\theta' =\alpha$, and the second at $\theta' =\beta$.
The points of intersection with the mirror are $(x_1,y_1)$ and $(x_2,y_2)$, respectively.
The difference in path lengths (lower path minus upper path) 
is $\Delta \ell = \Delta \ell_1 - \Delta \ell_2$, where
$\Delta \ell_2$ is illustrated.}
\label{fig:para2}
\end{figure}

The difference in the optical paths of these two rays ($\beta$ path minus $\alpha$
path) is denoted by $\Delta \ell$.
 The detailed expression for
this distance $\Delta \ell$ used in Refs.~\cite{FS00,FS02} is not quite correct, 
as was pointed out to us by Vuletic ~\cite{Vul}. The expression used in 
Refs.~\cite{FS00,FS02}, which we will denote by $\Delta \ell_1$, is the difference
in distance traveled by the two rays after they cross a line of constant $x$, 
perpendicular to the axis of the mirror. This difference is
\begin{equation}
\Delta \ell_1 = a\,|\cos\gamma (\cos\alpha -\cos\beta)
                  + \sin\gamma (\sin\alpha -\sin\beta)| \,.
\end{equation}
However, the difference in optical path lengths is the difference in distance travel-led
after crossing a line perpendicular to the incoming rays, as illustrated in 
Fig.~\ref{fig:para2}, and is
\begin{equation}
\Delta \ell = \Delta \ell_1 - \Delta \ell_2\,.
\end{equation}
The correction term, $ \Delta \ell_2$, is 
\begin{equation}
\Delta \ell_2 = a\,\left[ \sin \beta\, \sin(\beta-\gamma)
 -  \sin \alpha\, \sin(\alpha-\gamma) \right].
\end{equation}
The corrected expression for $\Delta \ell$ is then
\begin{equation}
\Delta \ell = a\,\left[ \cos\gamma (\cos\alpha -\cos\beta +\sin^2\alpha -\sin^2\beta)
+\sin\gamma (\sin\alpha -\sin\beta +\sin\beta \cos\beta -\sin\alpha \cos\alpha) \right]\,.
\end{equation}
The mean squared electric field near the focus of a parabola of revolution is,
in the geometric optic approximation,
\begin{equation}
\langle E^2 \rangle_{pr} = \frac{3 \hbar c}{2 \pi^2} \int \frac{d\theta}{(\Delta \ell)^4} \,, 
\label{eq:E2pr}
\end{equation}
The corresponding expression for a parabolic cylinder is
\begin{equation}
\langle E^2 \rangle_{pc} = \frac{16}{15 \pi}\,\langle E^2 \rangle_{pr} \,. 
\label{eq:E2pc}
\end{equation}
Note that in Eq.~(\ref{eq:E2pr}), the integration is over $\theta$, the angle of the
incident ray, not $\theta'$, the reflected angle, as was incorrectly stated in
Refs.~\cite{FS00,FS02}.

A detailed discussion of the electromagnetic case will be given elsewhere. Here
we are concerned with $\langle  \hat{\rho}^2 \rangle_R$ for the parabola of
revolution, which is obtained from
Eq.~(\ref{eq:E2pr}) by letting $c \rightarrow c_S$ and dividing by $2$, leading to the
result   
\begin{equation}
\langle  \hat{\rho}^2 \rangle_R =  \frac{3\hbar}{2 \pi^2\, c_S} 
\int_{\theta_{min}}^{\theta_{max}} \frac{d\theta}{(\Delta \ell)^4} \,,  \label{eq:near_focus}
\end{equation}
where the factor of $\hbar$ has been restored. The corresponding expression of a
parabolic cylinder is obtained by multiplying by ${16}/{15 \pi}$. Although the integrand
in the above expression is singular at $\Delta \ell =0$, it may be treated as a
distribution and the integral is well defined. 

\begin{figure}
\begin{center}
     \scalebox{0.3}{ 
\includegraphics{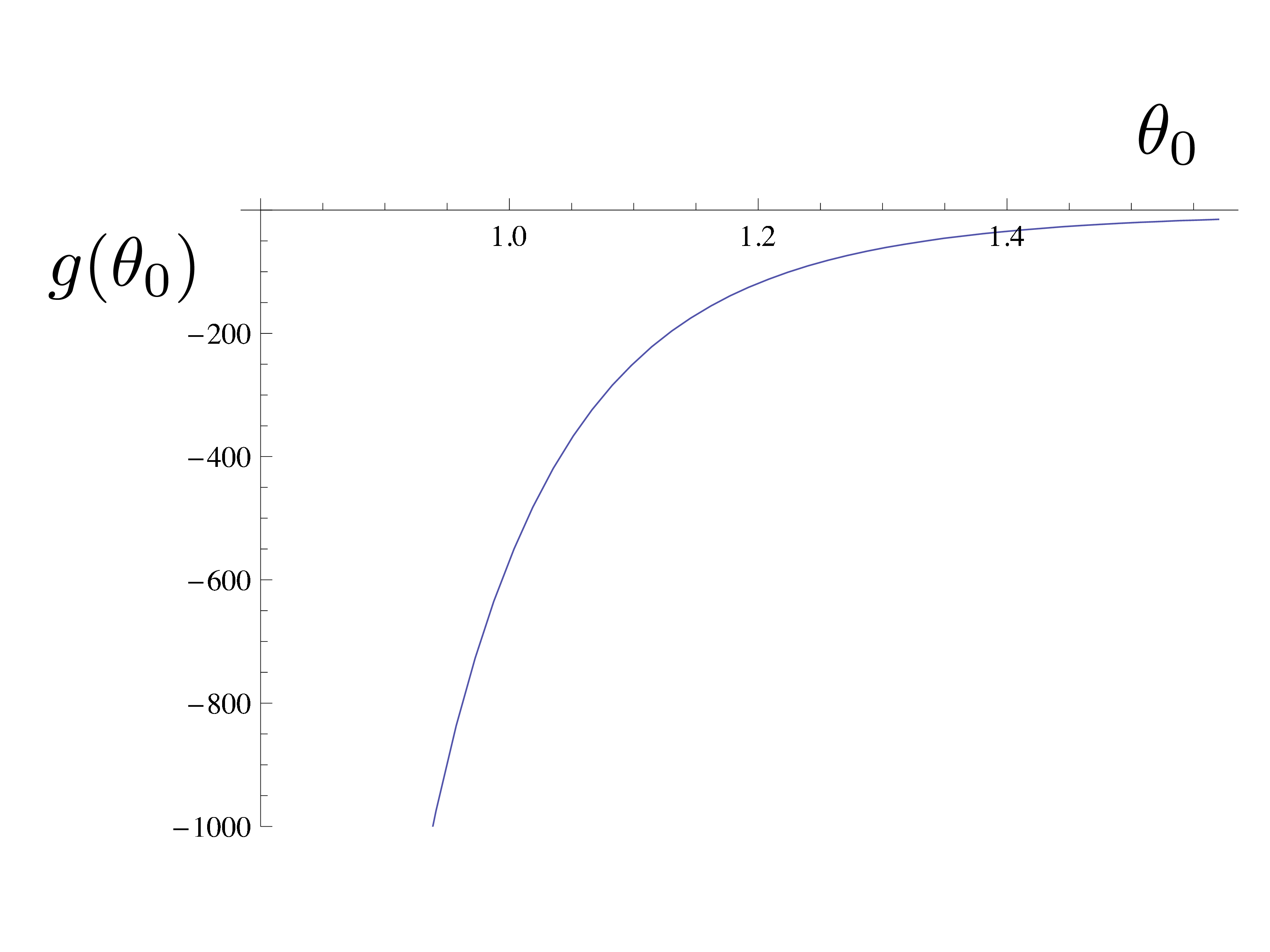}}
\end{center}
\caption{The function $g(\theta_0)$, defined in Eq.~(\ref{eq:g}), is plotted.}
\label{fig:g}
\end{figure}

Here we will treat only the case $\gamma = \pi/2$, where the integrations may be done
in closed form. In this case, $\beta = -\alpha$, as may be seen from the fact that
$f(\theta')$ is now an even function:
\begin{equation}
f(\theta') =  (1+\cos \theta')  \, \cos (\theta') \,. \label{eq:theta2}
\end{equation} 
The minimum value of $\theta$ in Eq.~(\ref{eq:near_focus}) is 
$\theta_{min}=(a/b)\,f(\theta_0)$, where $\theta_0$ is the angular size of the mirror.
The maximum value in our case is $\theta_{max}=2a/b$, corresponding to $\alpha =0$.
We have that 
\begin{equation}
\frac{d\theta}{d \alpha} = \frac{a}{b}\, f'(\alpha) =  
- \frac{a}{b}\,\sin \alpha \,(2\cos \alpha +1) \,.  \label{eq:theta/alpha}
\end{equation}
This relation may be used to express $\langle  \hat{\rho}^2 \rangle_R$ as
\begin{equation}
\langle  \hat{\rho}^2 \rangle_R =  -\frac{3\hbar}{32 \pi^2\, c_S\,a^3\,b} 
\int^{0}_{\theta_0} d\alpha\;\frac{\sin \alpha \,(2\cos \alpha +1)}{(\Delta \ell)^4} \,,
\end{equation}
or as
\begin{equation}
\langle  \hat{\rho}^2 \rangle_R =  \frac{3\hbar}{32 \pi^2\, c_S\,a^3\,b} 
\int_{0}^{\theta_0} d\alpha\;\frac{2\cos \alpha +1}{\sin^3 \alpha\, (1-\cos \alpha)^4}
 = \frac{3\hbar}{64 \pi^2\, c_S\,a^3\,b} \int_{-\theta_0}^{\theta_0} d\alpha\;
\frac{2\cos \alpha +1}{|\sin^3 \alpha|\, (1-\cos \alpha)^4} \,.
\end{equation}
 This integral may be performed explicitly, with the result
\begin{equation}
\langle  \hat{\rho}^2 \rangle_R =  \frac{3\hbar}{4096 \pi^2\, c_S\,a^3\,b} \; g(\theta_0)\, ,
\end{equation}
where
\begin{equation}
 g(\theta_0) = \log\left(\frac{1+\cos \theta_0}{1-\cos \theta_0} \right)
+\frac{30 \cos^5 \theta_0 -120\cos^4 \theta_0 +160\cos^3 \theta_0 -40\cos^2 \theta_0
-94\cos \theta_0 -224}{15 (1+\cos \theta_0) (1-\cos \theta_0)^5}\,. \label{eq:g}
\end{equation}
The function $g(\theta_0)$ is negative everywhere, and is plotted in Fig.~\ref{fig:g}.
The singularity as $\theta_0 \rightarrow 0$ represents a breakdown of the geometric
optics approximation, as diffraction effects become more important for small $\theta_0$.
For fixed $\theta_0$, the result is of the form
\begin{equation}
\langle  \hat{\rho}^2 \rangle_R = 
-\frac{ \hbar\, \rho_0\, C}{c_S \,b\, a^3} <0 \,, \label{eq:near_focus2}
\end{equation} 
where $C$ is a constant small compared to unity.

This, and the analogous expressions for $\langle E^2 \rangle$ and $\langle B^2 \rangle$,
which also are proportional to $1/(b\, a^3)$, are striking in that they can be large
when the focus is far from the mirror itself, $b \gg a$. This result is controversial,
and seems to be in conflict with a general result by Fewster and Pfenning~\cite{FP06},
which implies that quantities such as $\langle E^2 \rangle$ or 
$\langle  \hat{\rho}^2 \rangle_R$
should be proportional to
the inverse fourth power of the distance to the mirror, which is to say 
$\propto b^{-4}$ in this case. On the other hand, there is a simple physical argument
to the contrary, which we find compelling: 
the interference term between multiply reflected rays
is slowly oscillating when $\Delta \ell \propto a$ is small, and should give a 
contribution proportional to an inverse power of $a$, as in Eq.~(\ref{eq:near_focus2}). 
In any case, the study of the phononic case provides an additional theoretical,
and potentially experimental, probe to better understand this issue.

\section{Summary and Discussion}
\label{sec:final}

In this paper, we have treated the effects of squeezed phonon states and of boundaries
on the local quantum density fluctuations of a fluid, assuming a linear phonon
dispersion relation. The purpose of this investigation is two-fold. The modified
density fluctuations are of interest in their own right and are in principle
observable by light or neutron scattering. Secondly, the phononic system studied
here is a potentially useful analog model for better understanding quantum fluctuations
in relativistic quantum field theory with boundaries. After reviewing the density
fluctuations in a boundaryless system in the phonon vacuum state, we treated the
effects of a squeezed vacuum state of phonons. Here we found that such a state
will have both local increases and local decreases in the mean squared density. However,
the time or spatial averaged effect is an increase. This is in complete analogy
to the case in relativistic quantum field theory, with the decrease in mean squared density
corresponding to regions of negative energy density. 

We next turned our attention to the effects of perfectly reflecting boundaries and 
studied the cases of one and two parallel plates, a torus, a wedge, a cosmic string,
and a parabolic mirror. In all of the cases examined, we found a decrease in
mean squared density, $\langle  \hat{\rho}^2 \rangle_R <0$. This amounts to a suppression
of the usual zero point fluctuations, and is analogous to the suppression of vacuum
fluctuations which can lead to negative energy density in quantum field theory. In
general, $\langle   \hat{\rho}^2 \rangle_R$ due to boundaries is inversely proportional
to the speed of sound, $c_S$. This is in contrast to total energies or forces, such as
Eq.~(\ref{eq:force}), which are proportional to $c_S$, and to the mean squared electric 
or magnetic fields near a perfect reflector, Eqs.~(\ref{eq:E2}) and (\ref{eq:B2}),
which are proportional to the speed of light.

The case of the parabolic mirror is of particular interest. Here we were able to 
correct certain aspects of our previous treatment~\cite{FS00,FS02} for 
electromagnetic fields. We
find that near the focus,  $\langle  \hat{\rho}^2 \rangle_R$ grows as the inverse
cube of the distance ot the focus. For the phononic case, this growth necessarily
stops as the scale of interatomic spacing is reached. However, the analysis performed
here for phonons also applies to the case of the quantized electromagnetic field,
where one expects the same rate of growth in the mean squared electric and magnetic 
fields.

\vspace{1cm}

\begin{acknowledgments}
This work was supported in part by the National Science Foundation
under Grant PHY-0555754 and by Conselho Nacional de Desenvolvimento
Cientifico e Tecnologico do Brasil (CNPq). LHF would like to thank the Institute
of Physics at Academia Sinica in Taipei and National Dong Hwa University in
Hualien, Taiwan for hospitality while this manuscript was completed.
\end{acknowledgments}


\begin{thebibliography}{40}

\bibitem{DLP} I.E. Dzyaloshinskii, E.M. Lifshitz and L.P. Pitaevski,
Adv. Phys. {\bf 10}, 165 (1961).

\bibitem{Larraza} A. Larraza, Phys. Lett. A {\bf 248}, 151 (1998).

\bibitem{Bschorr} O. Bschorr, J. Acoust. Soc. Am. {\bf 106}, 3730 (1999).

\bibitem{SU02} E. Sch{\"a}ffer and U. Steiner, Eur. Phys. J. E {\bf 8},
347 (2002).

\bibitem{RP05} D.C. Roberts and Y. Pomeau, Phys. Rev. Lett. {\bf 95},
145303 (2005).

\bibitem{Recati} A. Recati, J.N. Fuchs, C.S. Pe{\c c}a, and
  W. Zwerger, Phys. Rev. A {\bf 72}, 023616 (2005).

\bibitem{Lamoreaux} S.K. Lamoreaux, arxiv:0808.4000.

\bibitem{Unruh}  W. G. Unruh Phys. Rev. Lett. {\bf 46 }, 1351 (1981); 
Phys. Rev. D {\bf 51}, 2827 (1995)  

\bibitem{FS08}  L.H. Ford and N.F. Svaiter, arXiv:0811.2409.

\bibitem{LL-ST} See, for example, E.M. Lifshitz and  L.P. Pitaevski,
{\it Statistical Physics, Part 2}, 2nd ed. (Pergomon, Oxford, 1969),
Eq.~(24.10).

\bibitem{FF04} P.O. Fedichev and U.R. Fischer, Phys. Rev. A {\bf 69}, 033602
(2004).

\bibitem{FS09}  L.H. Ford and N.F. Svaiter, Phys. Rev. Letts. {\bf 102}, 030602 (2009).

\bibitem{footnote} Note that the thermal Brillouin scattering cross section cited in 
Eq.~(25) of Ref.~\cite{FS09} is actually the high temperature limit of the total 
Brillouin cross section and hence twice the Stokes line cross section. This compensates
for the fact that $1/2$ of the zero point part is canceled by the thermal correction.
Thus the ratio $R$ in Eq.~(27) of Ref.~\cite{FS09} is still correct.  

\bibitem{Fetter} A.L. Fetter, J. Low Temp. Phys. {\bf 6}, 487 (1972).

\bibitem{FF56} R.P. Feynman and M. Cohen, Phys. Rev. {\bf 102}, 1189 (1956).

\bibitem{Caves}  C. Caves, Phys. Rev. D {\bf 23}, 1693 (1981).

\bibitem{GC08} J. C. Garisson and R. Y. Chiao, {\it Quantum Optics"}
(Oxford University Press, Oxford, 2008).

\bibitem{KF93} C.-I Kuo and L.H. Ford, Phys. Rev. D {\bf 47}, 4510 (1993).

\bibitem{BFR02}  A. Borde, L.H. Ford, and T.A. Roman,  Phys. Rev. D {\bf 65} 084002 
(2002).

\bibitem{Jaffe} R. L. Jaffe,  arXiv:hep-th/0307014.

\bibitem{MCW06}  K. A. Milton, I. Cavero-Pelaez, and J. Wagner J. Phys. A {\bf 39},
6543 (2006).  (Preprint  hep-th/0510236)

\bibitem{SF05}  V. Sopova and  L.H. Ford, Phys. Rev. D {\bf 72}, 105010 (2005)
 (Preprint  quant-ph/0504143)

\bibitem{FS98}  L.H. Ford and N.F. Svaiter,  Phys. Rev. D {\bf 58}, 065007 (1998) 
 (Preprint  quant-ph/9804056)

\bibitem{RS03} R. B. Rodrigues and N. F. Svaiter, Physica A {\bf 328}, 466 (2003).

\bibitem{CD79} P. Candelas and D. Deutsch, Phys. Rev. D {\bf 20}, 3063 (1979). 

\bibitem{HK86} T.M. Helliwell and D.A. Konkowski, Phys. Rev. D {\bf 34}, 1918 (1986).

\bibitem{FS00}  L.H. Ford and N.F. Svaiter Phys. Rev. A {\bf 62}, 062105 (2000)
 (Preprint  quant-ph/0003129)

\bibitem{FS02}  L.H. Ford and N.F. Svaiter,  Phys. Rev. A {\bf 66}, 062106 (2002)
 (Preprint  quant-ph/0204126)

\bibitem{Vul} V. Vuletic, private communication.

\bibitem{FP06} C.J. Fewster and M.J. Pfenning, J. Math. Phys. {\bf 47}, 082303 (2006) 



\end{thebibliography}
\end{document}